\documentclass[]{spie}  

\usepackage{amsmath,amsfonts,amssymb}
\usepackage{graphicx}
\usepackage{svg}
\usepackage[colorlinks=true, allcolors=blue]{hyperref}

\newcommand{\farcs}{$\,.\!\!^{\prime\prime}$}
\newcommand{\arcsec}{$^{\prime\prime}$}
\newcommand{\arcmin}{$^{\prime}$}
\newcommand{\de}{$^{\circ}$\,}
\newcommand{\rotpposn}{\texttt{rotpposn}}
\newcommand{\Th}{$^{th}$}
\newcommand{\rd}{$^{rd}$}

\newcommand{\note}[1]{{#1}} 
\newcommand{\sref}{\S\ref} 

\title{Flexure updates to MOSFIRE on the Keck I telescope}

\author[a]{Taylor A.\ Hutchison\,*\,}
\author[b]{Josh Walawender}
\author[b]{Shui Hung Kwok}
\affil[a]{Mitchell Institute for Fundamental Physics \& Astronomy, Texas A\&M University, 4242 TAMU, College Station, TX, 77843, USA}
\affil[b]{W.\ M.\ Keck Observatory, 65-1120 Mamalahoa Hwy, Waimea, HI, 96743, USA}

\authorinfo{* NSF Graduate Fellow \\ Contact: T.A.H., aibhleog@tamu.edu}

\begin{document} 
\maketitle

\begin{abstract}
We present a recent evaluation and updates applied to the Multi-Object Spectrometer For Infra-Red Exploration (MOSFIRE) \cite{mcle12} on the Keck I telescope. Over the course of significantly long integrations, when MOSFIRE sits on one mask for $>$4 hours, a slight drift in mask stars has been measured. While this does not affect all science-cases done with MOSFIRE, the drift can smear out signal for observers whose science objective depends upon lengthy integrations. This effect was determined to be the possible result of three factors: the internal flexure compensation system (FCS), the guider camera flexure system, and/or the differential atmospheric refraction (DAR) corrections. In this work, we will summarize the three systems and walk through the current testing done to narrow down the possible culprit of this drift and highlight future testing to be done.
\end{abstract}

\keywords{Keck, MOSFIRE, spectroscopy, near-infrared, flexure, drift} 

\section{Introduction} \label{sec:intro} 
The Multi-Object Spectrometer For Infra-Red Exploration (MOSFIRE) on the Keck I telescope has been the driver of countless scientific advances in astronomy.  The unparalleled sensitivity and multiplexing of MOSFIRE has made it a coveted instrument across the general astronomy community.  Just one year after commissioning (c.\ 2012 \cite{mcle12}), it heralded the detection of the most distant galaxy ever observed at the time ($z=7.5078$, only 800 Myr after the Big Bang \cite{fink13}).  Since then, it has enabled the detailed study of celestial objects both near and far -- 
from small and variable brown dwarfs \cite{manj20} 
to the young, massive stellar populations in the most distant sources \cite{hutc19}.

A few years after commissioning, detailed studies of data taken using MOSFIRE (when observers remained the same mask for more than two hours) discovered a slight spatial drift of 1 pix/hr in the reduced spectra. This drift was carefully measured and identified as a combination of the flexure between the optical guider and near-infrared (NIR) science fields of view (FOVs) as well as the effects from differential atmospheric refraction (DAR) \cite{kass15}.  The effects were corrected for via modeling the drift as a function of elevation and rotation. 

More recently, some MOSFIRE users have discovered that a very slight drift still remains in the data, however this effect is only noticeable in observations where observers use the same mask continuously for $>$4 hours.  The method to track this drift is straightforward: 1) place an alignment star on a slit in the mask and measure the peak of the raw spatial profile for the star, then 2) measure the spatial peak location for the alignment star over the course of the observation, and 3) compare to the first frame's location to measure the drift of the target mask.  From this process, observers measured a spatial drift of about 0.5 pix/hr. 

\begin{figure}[ht]
    \centering
    \includegraphics[trim=0 25 0 35,clip,width=\linewidth]{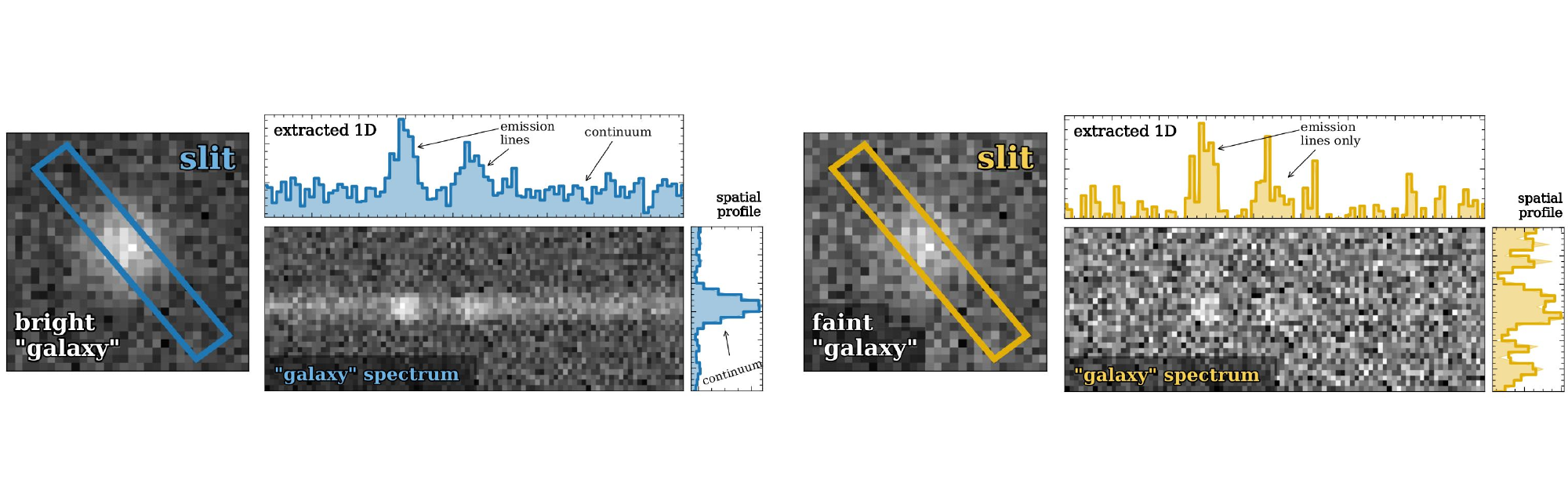}
    \caption{Simulated spectra for two fake galaxies (no drift added). (\textit{left}) For a reasonably bright source ($H_{160} < 25$ mag), $>$4 hrs produces detections of both strong emission lines and the source's continuum.  Both of these features have been identified in the extracted 1D spectrum and the collapsed spatial profile. (\textit{right}) For a fainter source ($H_{160} > 25$ mag), $>$4 hrs may result in a faint detection of just the emission lines (not enough to detect continuum). The faint emission lines have been identified in the extracted 1D spectrum.}
    \label{fig:sim-galaxies}
\end{figure}

This smaller drift is still a concern. While 0.5 pix/hr would not impact most MOSFIRE users, it becomes crucial for those who have MOSFIRE sit on one field for more than four hours. The subtleties of this effect may not be immediately obvious for those not immersed in these kind of observations, especially for such faint objects \& their emission lines.  Figure \ref{fig:sim-galaxies} shows the differences in spectroscopy for a reasonably bright galaxy ($H_{160} < 25$ mag), where both strong emission lines and continuum can be seen, versus a faint galaxy ($H_{160} > 25$ mag), where only faint emission lines are found.  Even with very long integrations, one would never expect to detect continuum for these significantly fainter sources, as it would likely always fall within the noise threshold.

\begin{figure}[ht]
    \centering
    \includegraphics[trim=0 25 0 30,clip,width=\linewidth]{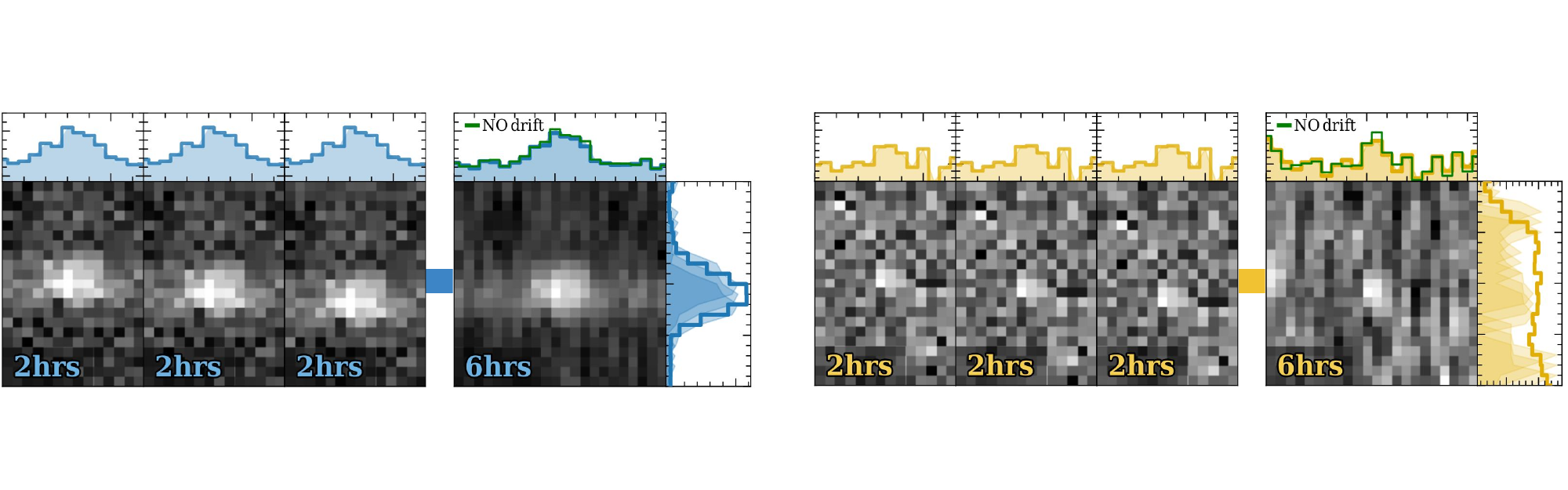}
    \caption{Simulated spectra for two fake galaxies with a simulated drift of 0.5 pix/hr added. The resulting stacked spectrum and extracted spectrum are on the right of each plot (with a green line overlaid showing the strength of the emission line when there is no drift). (\textit{left}) For the brighter galaxy ($H_{160} < 25$ mag), while the shift can been seen in the gaussians in the final stack's spatial profile, the overall final profile shape is hardly changed. (\textit{right}) For the fainter galaxy ($H_{160} > 25$ mag), the effect from the drift is more noticeable and significant when working with faint emission lines, as can be seen in the increase in flux gained from the extracted 1D spectrum when there is no drift (green line).}
    \label{fig:sim-drift}
\end{figure}

The effects of drift on the two different sources, over long integrations, can be very disparate, especially if the source is faint enough that one requires many hours of integration to gain enough signal-to-noise to see emission lines at all.  Figure \ref{fig:sim-drift} shows an example of the effect created by this 0.5 pix/hr drift for both a bright source (with continuum) versus a faint source (only faint emission lines).  For both panels, the simulated galaxy is shifted spatially 0.5 pix/hr for each 2 hr chunk, with a final stacked 2D spectrum shown for the total 6 hrs.

For a bright source, this effect does not necessarily pose a large problem; but for fainter sources, where the emission feature may span only a few pixels spatially (see the right panel of Figure \ref{fig:sim-drift} for this example), this drift can spread out the emission feature enough that it can become undetectable or incredibly challenging to locate. Adding to this problem, the continuum or emission lines for very faint sources ($H_{160} > 25$ mag) are never seen in the raw MOSFIRE data, therefore to identify even a potential galaxy detection in these datasets would require data reduction and stacking.  This becomes even more important to characterize when emission lines are so faint that neither they nor the galaxy's continuum can be seen in a few hours of data, therefore making it challenging to know how to shift and stack the data using modified data reduction pipelines.  One can see how this process can easily become disrupted by a slight drift smearing out the potential signal of these fainter galaxies, leading possibly to more claims of ``non-detections'' than are truly warranted.

In this manuscript, we will outline the current testing and analysis done to investigate this remaining spatial drift.  We will highlight the systems we suspect are responsible and discuss the ways these system could contribute to the observed drift.  
The three possible culprits of this spatial drift are described in \sref{sec:culprits}.  The initial testing done and current results are described in \sref{sec:testing}.  Finally, discussion about the results and future work planned are in \sref{sec:future}.

\newpage 

\section{The MOSFIRE System \& the source of the drift} \label{sec:culprits}

The Multi-Object Spectrometer For Infra-Red Exploration (MOSFIRE) is a NIR (0.97$-$2.41 $\mu$m) multi-object spectrometer and imager, with a science detector FOV of 6.1\arcmin\ x 6.1\arcmin\ and a spectroscopic resolving power of R$\sim$3,500 for a 0\farcs7 slit. The spectrograph is equipped with four NIR order-sorting filters $Y$, $J$, $H$, and $K$ (for the 6\Th, 5\Th, 4\Th, and 3\rd\ orders, respectively).\footnote{The $K$ filter cannot be used for imaging, however there is a $K_S$ filter that can be substituted. Additionally, there are $J2, J3, H1,$ and $H2$ filters available for imaging.}
MOSFIRE was built with a cryogenic Configurable Slit Unit (CSU) which can be reconfigured into a new mask while on sky in less than five minutes, with the ability to take spectroscopy of up to 46 objects simultaneously.

\begin{figure}[ht] 
    \centering
    \includegraphics[width=0.8\linewidth]{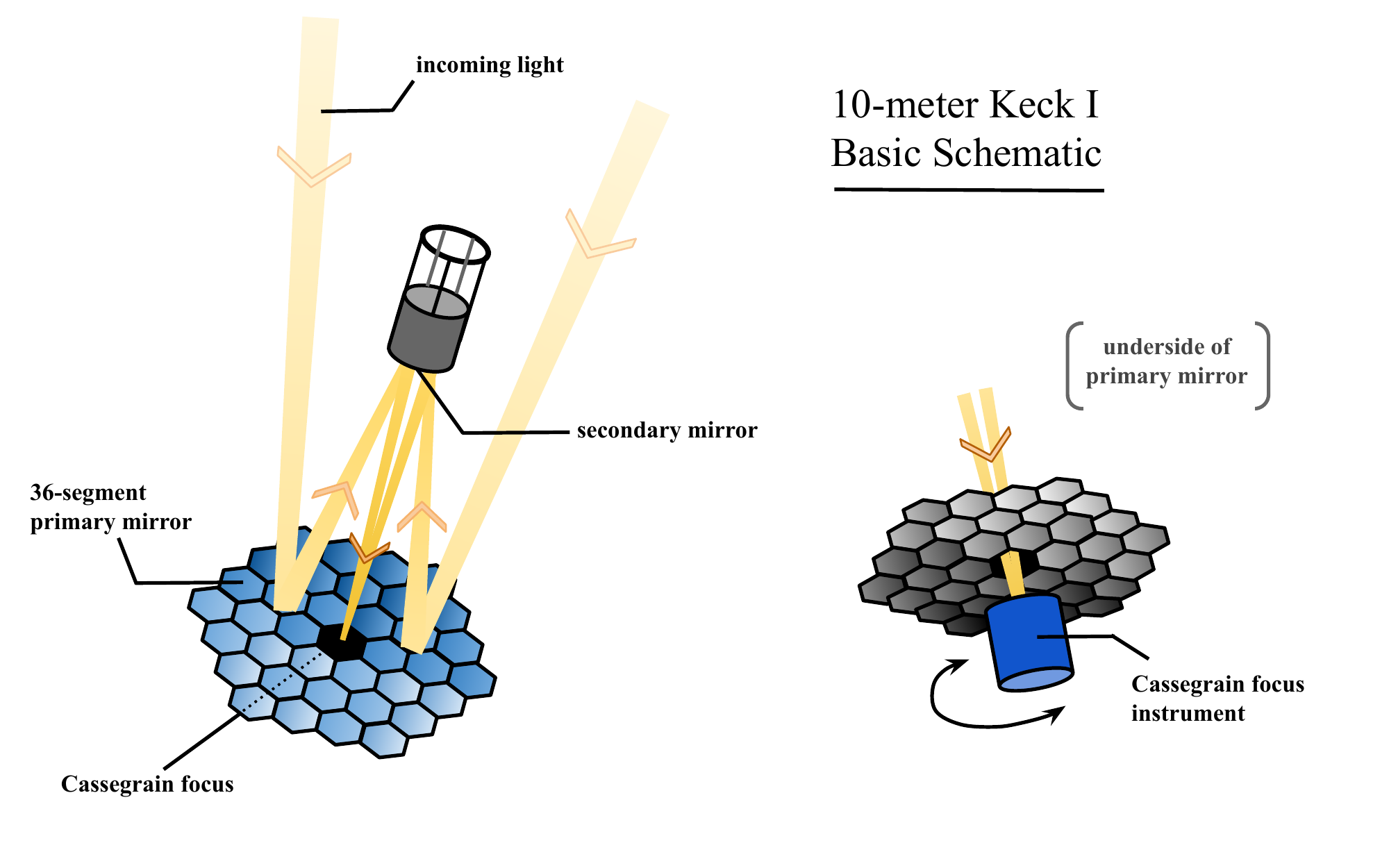}
    \caption{Basic schematic of the Keck I telescope, showing the path of the incoming light and the locations of the primary and secondary mirrors.  Additionally, the underside of the primary mirror is shown, highlighting the location of Cassegrain instruments such as the MOSFIRE spectrograph. The curved arrow indicates the rotation of the instrument. \label{fig:keck-schematic}}
\end{figure}

The MOSFIRE spectrograph is located at the Cassegrain focus of the telescope (see Figure \ref{fig:keck-schematic}).  As a direct result of being in this location, the instrument is subject to gravity pulling on it in different orientations as the telescope moves around in elevation and rotation while tracking cosmic objects through the night.  Therefore, the gravity vector pulling on the instrument changes as the telescope orientation changes.

During construction, the MOSFIRE team\cite{mcle12,mcle10} extensively tested and corrected for the expected flexure to be experienced by the instrument, however in use more correction was found to be necessary\cite{kass15} (as described briefly in \sref{sec:intro}). In general, there are only three main culprits for this measured remaining drift:

\begin{list}{--}{}
    \item Internal Flexure Compensation System (FCS)
    \item Optical Guider Flexure System
    \item Differential Atmospheric Refraction (DAR)
\end{list}

In this section, we will dive into detail about the three different culprits and how they are integrated into the entire telescope\,$+$\,instrument system.

\newpage
\subsection{Internal Flexure Compensation System} \label{ssec:fcs}
The Internal Flexure Compensation System (FCS) is a mirror located within the optical path of the MOSFIRE spectrograph whose purpose is to correct for warping of the instrument due to the gravity vector pulling on the spectrograph.  This warping results in the images of the slitmask moving on the detector pixels as the instrument orientation changes.

The FCS accounts for this gravity vector changing in two steps: (1) a pixel shift is calculated based upon the instrument's ``attitude'', and (2) the pixel shifts are converted into angle shifts on the FCS (which is implemented via changing piston values).  The changing piston values control the piezoelectric actuators, which tip and tilt the FCS mirror to correct for the resulting flexure.  The entire piezoelectric tip/tilt platform was designed and manufactured by Physik Instrumente \cite{mcle12}.

\begin{figure}[ht]
\begin{minipage}{0.53\textwidth}
    \centering
    \includegraphics[width=\linewidth]{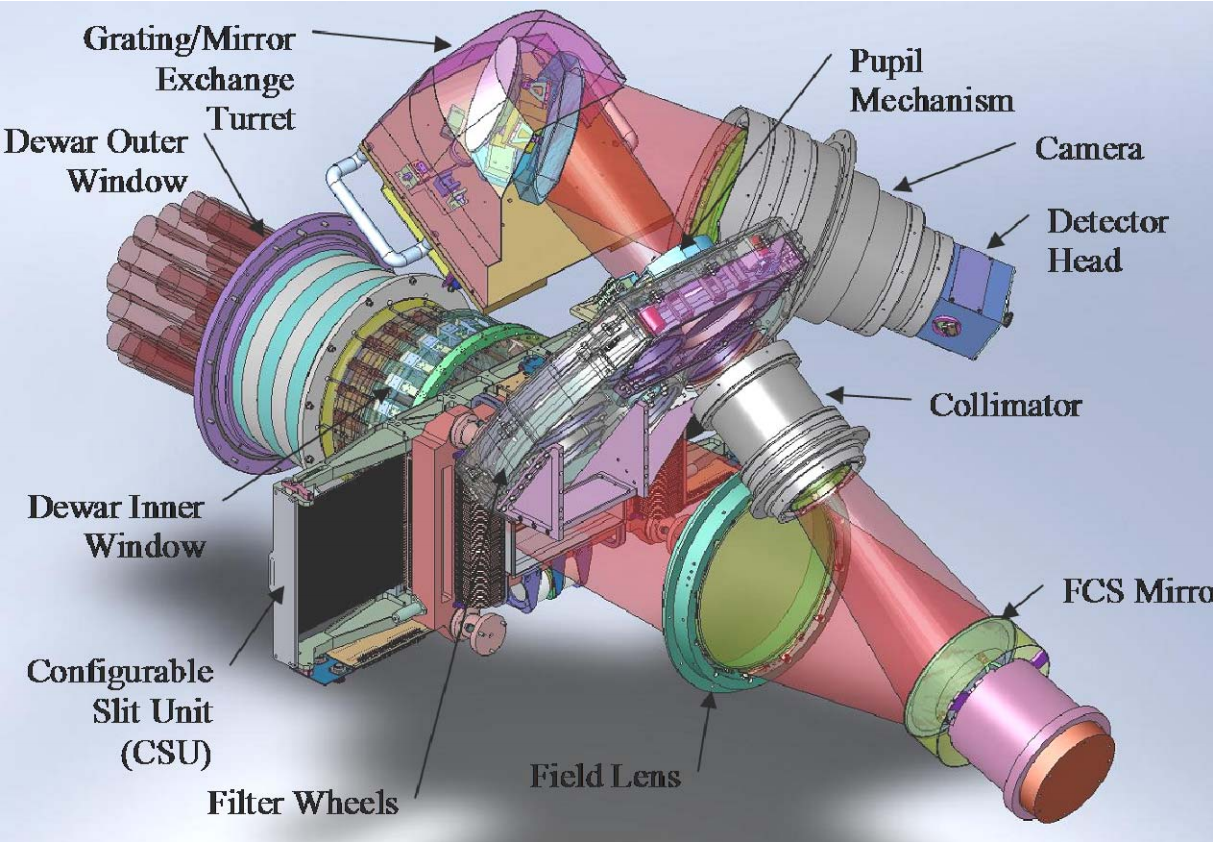}
    \caption{\label{fig:optical-path}Full optical train for MOSFIRE, from Ref. \citenum{mcle12}}
\end{minipage}
\begin{minipage}{0.3\textwidth}
\end{minipage}
\begin{minipage}{0.45\textwidth}
    \centering 
    \includegraphics[width=\linewidth]{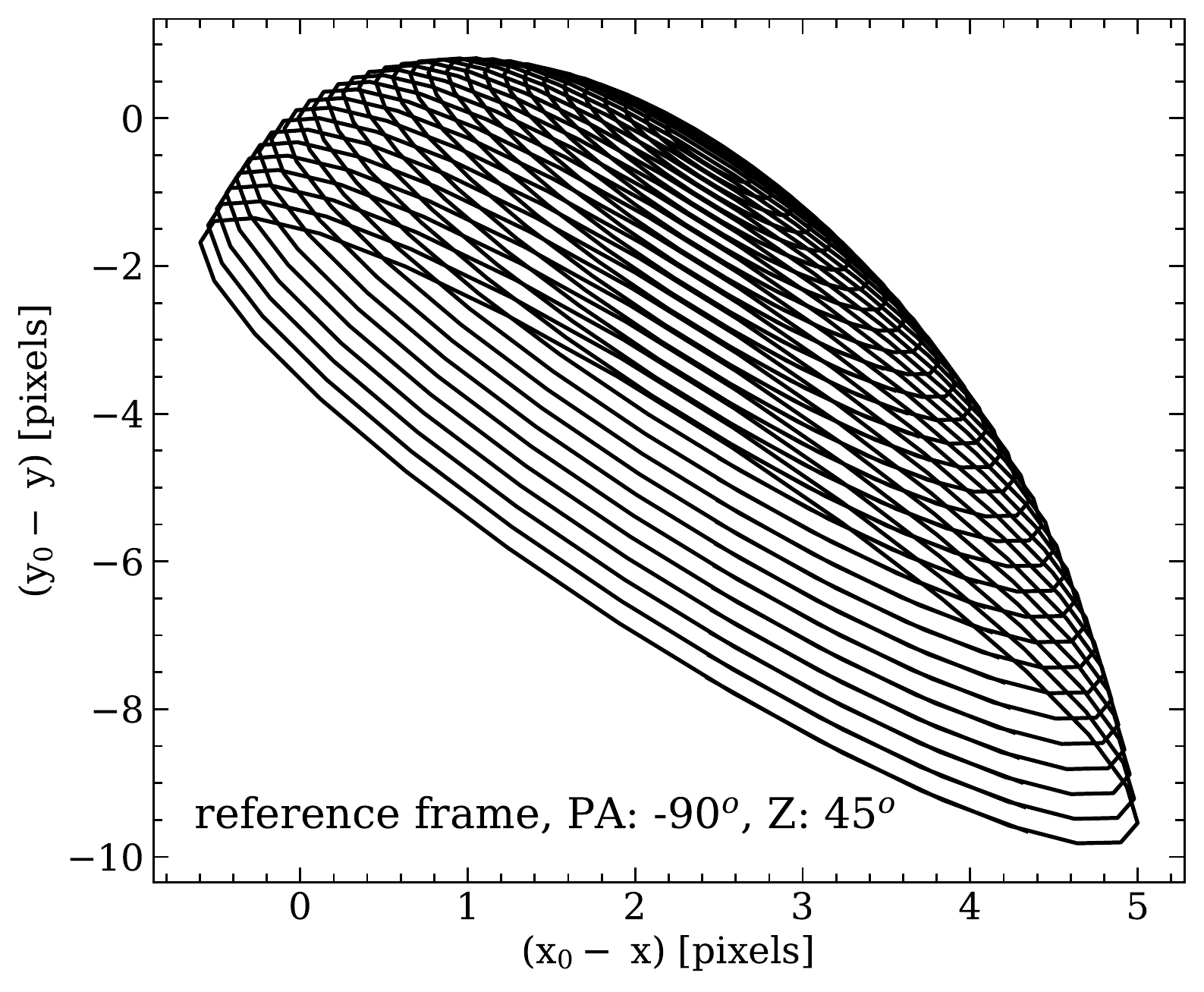}
    \caption{\label{fig:fcs-model}FCS model pixel shifts for a range of elevation and physical rotator angle compared to a reference.}
\end{minipage}
\end{figure}

Figure \ref{fig:optical-path} shows a diagram of the full optical path for the MOSFIRE spectrograph, with the FCS mirror located on the bottom right.  Currently, the FCS is described by a model which calculates the pixel shift, using measured parameters provided by a given coordinate system and lookup table.
Figure \ref{fig:fcs-model} shows the way this model works compared against a reference frame of elevation = 45\de ($90^{\circ}{-}Z$=EL) and physical rotator angle = -90\de (the same reference frame will be used in our actual measurements in \sref{ssec:test-fcs} for easier comparison), running through a range of elevations and physical rotator angles (\rotpposn).  To see an animated version of this, view the GitHub repository\footnote{GitHub repository, \href{https://github.com/aibhleog/Keck-Visiting-Scholar}{github.com/aibhleog/Keck-Visiting-Scholar}} associated with this work and navigate \href{https://github.com/aibhleog/Keck-Visiting-Scholar/blob/master/plots-data/data_FCS/shifts_with_reference.gif}{here}.

\subsection{Guider Flexure System}
\begin{figure}[ht]
    \centering
    \includegraphics[width=0.75\textwidth]{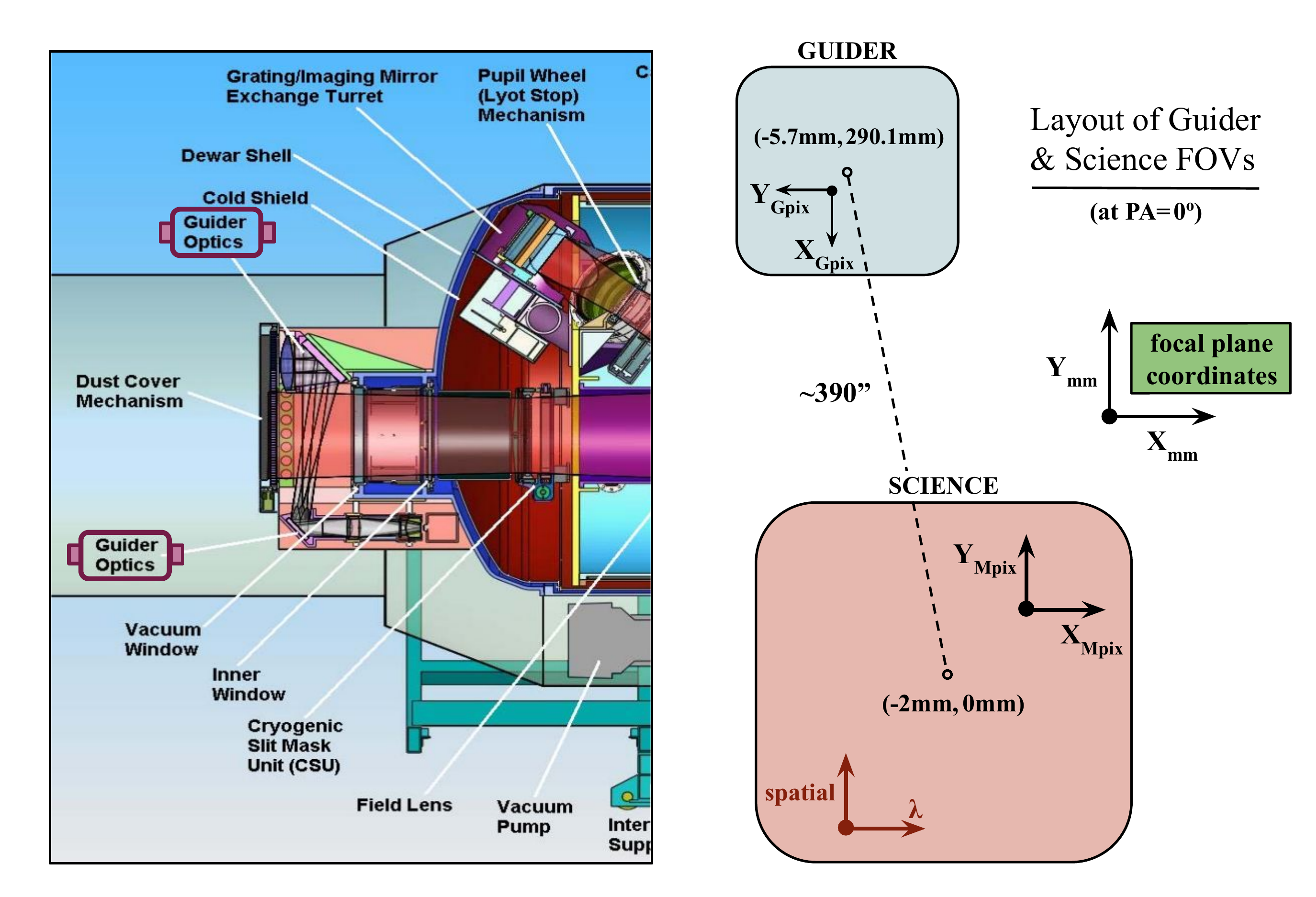}
    \caption{(\textit{left}) Optical path of MOSFIRE from Ref. \citenum{mcle10}, zoomed in to highlight the location of the guider and the guider optics involved.  The ``guider optics'' tags in the diagram are encircled with purple rectangles. (\textit{right}) Layout of the guider and science detector FOVs, along with their respective coordinate systems and their separation on sky at PA = 0\de\!.}
    \label{fig:guider-optics}
    ~\vspace{-5mm}\\
\end{figure}

As the telescope tracks through the night, the guider camera observes a guide star to keep the science targets in their exact slit locations over the course of the observations.  The guider camera is mounted as an independent off-axis CCD \cite{mcle10} (see the left panel of Figure \ref{fig:guider-optics} for the guider location and guider optics, tags highlighted by purple rectangles). The guider operates in the optical wavelengths, and has a remote focus control \cite{mcle12}. The guider control is run by the Keck Observatory's standard MAGIQ software.  

Similar to the internal flexure, there is a flexure component between the guider camera and the science detector that needs to be corrected. \note{This flexure is compensated for in software by taking the measured ($x$,$y$) location of the guide star in the guider FOV and adding a correction using elevation and rotation as inputs (applied via a sinusoidal combination).  This results in moving the pixel location of the guide star over time as a function of elevation and rotation, such that the image of the CSU remains in the correct location on the science detector.}

The right panel of Figure \ref{fig:guider-optics} (not to scale) shows the FOV for the science detector (where the CSU is located) and the location and offset angle where the guider camera is located.  As shown, the center of the science detector's FOV is approximately 390\arcsec\ away from the guider's FOV. This means that as the telescope changes in elevation and rotation throughout the night, the distance and tension between the two FOVs change -- resulting in the image of the CSU on the science detector to move.  This effect was originally corrected for in the instrument design \cite{mcle12}, however in the investigation done 
in 2015 \cite{kass15} they found that more refinement was necessary.

\subsection{Differential Atmospheric Refraction}
The differential atmospheric refraction (DAR) is the effect on light coming from a distant object being refracted by the atmosphere.  The effects of this can be significant, where the location of a star can appear to change between two different bandpasses (filters), and is increasingly important to characterize as observers look to redder wavelengths.  Additionally, the closer an object is to the horizon, the more severe this effect will become -- as blue light is scattered away at higher rates than red light.

For MOSFIRE, the DAR is taken into account for the guiding system and how its location on the observable sky relates to the science FOV.  The MOSFIRE guider operates in the optical wavelengths, while the science detector operates in the NIR.  Therefore, the location of the guiding star may be slightly different in the NIR than in the optical -- thus, a small additional adjustment must be made to account for this while guiding, such that the targets remain in the correct locations in their respective slits.  The current system is set up to account for all of these factors.

\section{Initial testing \& preliminary results} \label{sec:testing}

The initial testing done is described in this section.  Firstly, we will discuss insights derived from existing data -- spectroscopic data taken by MOSFIRE observers, where at least one alignment star had its own slit.  Secondly, we will describe the initial testing done to track down the source of the remaining drift, following the order of potential culprits listed in \sref{sec:culprits}.
 
\subsection{Insights from Existing Data} \label{ssec:insights}

This remaining drift has been mentioned in several studies, both in large surveys (e.g., the MOSDEF team \cite{krei15}) and in small programs which stared at the same targets for over six hours (e.g., I. Jung \cite{jung20} \& R. Larson \cite{lars20}).  There are two ways we can gain insight into this remaining drift from existing data.  First, using data with an alignment star that has its own slit, we can track the relative drift of the star (spatially) in the data throughout the night.  Any measured drift here would suggest that at least some part of the remaining drift is due to the flexure between the optical guider and the science detector or uncorrected DAR.  Second, running cross-correlations between raw frames (for the total science mask, but with detectable continuum masked out), we can track the relative shift of the slits themselves based upon the atmospheric OH lines (``skylines'') in the NIR that would be present in these data.  Any measured shifts here would suggest that at least some part of the remaining drift is due to incomplete correction of internal flexure by the FCS.

The following sections will detail these processes and share the results.

\begin{figure}[hbt]
    \centering
    \includegraphics[width=\linewidth]{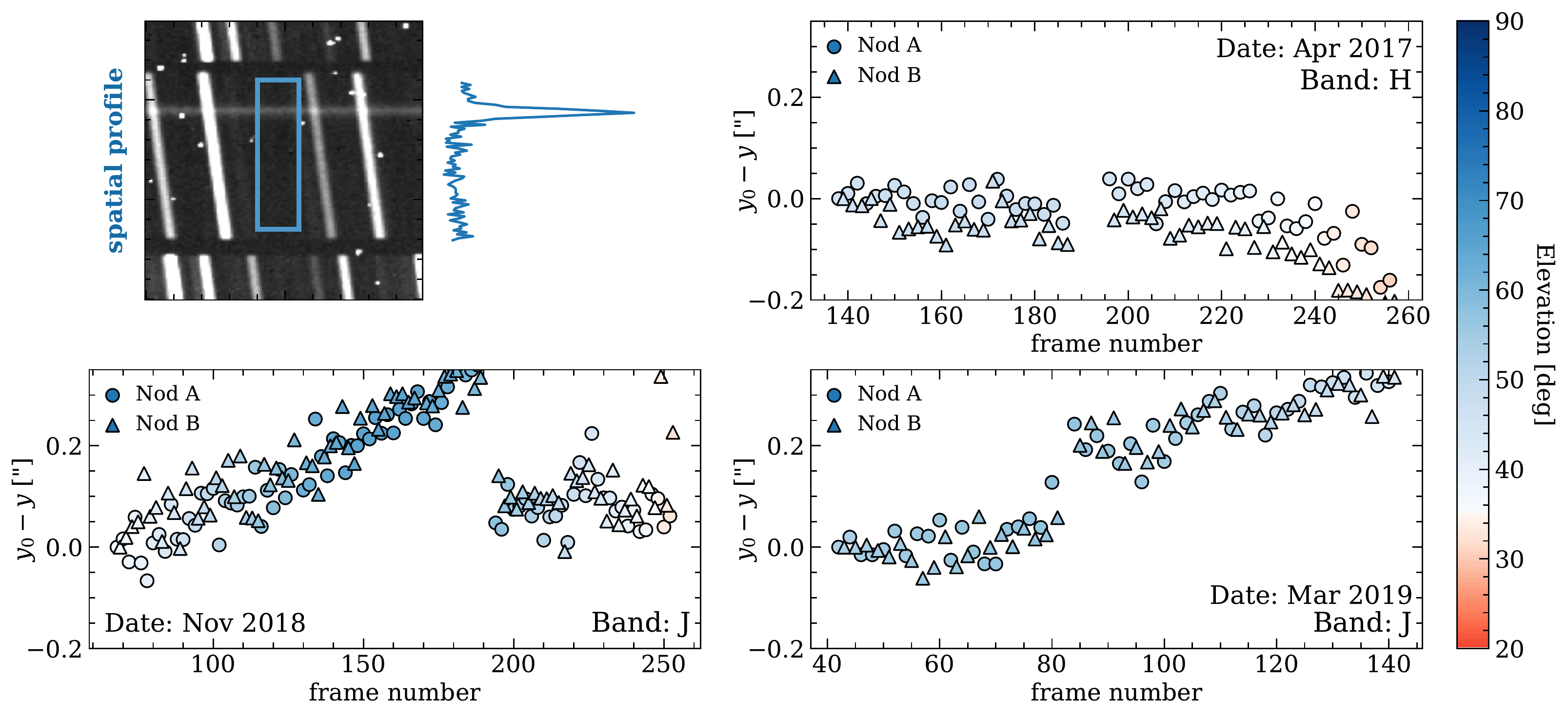}
    \caption{(\textit{top left}) An example of a raw $J$-band frame, zoomed in on the slit holding an alignment star.  The blue rectangle marks a region devoid of atmospheric skylines that is used to generate the spatial profile seen in the right of the 2D spectrum.  This profile can then be used as the measurement of the star's spatial location in that raw frame. (\textit{top right \& bottom}) Three example datasets from $2017-2019$ showing different examples of the relative spatial drifts of an alignment star in these datasets.  The scatter points are colored to indicate elevation for each frame, and the scatter point shapes denote the two different nods used in observing. Note that the pixel scale is 0\farcs18/pix.}
    \label{fig:star-drift}
\end{figure}

\subsubsection{Checking the Target's Drift} \label{sssec:insights-star}
The method for tracking this drift over the course of an observation has been briefly described in \sref{sec:intro}, however we will list it again here (and in more detail) for clarity: \\

\hspace{3mm} \textbf{Measuring the Relative Offset (Drift) from Raw Data:}
~\vspace{-7mm} \\
\begin{enumerate}
    \item place one or more alignment stars on their own slits in the MOSFIRE mask design 
    \item measure the collapsed spatial profile of the star in each raw MOSFIRE frame 
    \item compare the spatial location of the alignment star (using the collapsed spatial profile) in each frame to its spatial location in the first frame
\end{enumerate}

There are some subtleties to this process that will need to be accounted for, such as the fact that the raw frames are not rectified and therefore there will be a slight curve to the shape of the star's spectrum. Additionally, as is prevalent in the NIR, there will be a slew of atmospheric skylines which will add noise to a collapsed spatial profile.  To account for this, we recommend collapsing only a small slice of the raw spectrum in an area devoid of skylines, such that the spatial profile clearly shows the gaussian shape of the star's continuum (the top left of Figure \ref{fig:star-drift} shows what this small slice of the raw spectrum could look like).

From this process, we can measure the drift of an object over the course of the entire observation.  Figure \ref{fig:star-drift} shows measurements of this drift for a handful of datasets -- note that these datasets were chosen to be after 2015, as that was when the most recent updates to the flexure models and DAR corrections were applied.  In the different panels, the scatter points (raw frames) are colored based upon the elevation at the time of the exposure. Additionally, the two different scatter point shapes indicate that these data were all taken using the nod technique, where the telescope keeps the target in the slit but nods back and forth to allow different background sky to come through.  This is a common technique used to allow observers the ability to remain on target the entire time (versus moving off target after every exposure to take background frames).

For the various datasets shown in Figure \ref{fig:star-drift}, there is clearly a drift measured.  In some cases, the drift is more or less severe, but nevertheless it exists.  Therefore, we infer that some fraction of the remaining drift in MOSFIRE must be due to the flexure between the optical guider and the science detector.  We will outline the further testing done to investigate this in \sref{ssec:test-guider}.

\subsubsection{Checking the Science Mask's Drift} \label{sssec:insights-mask}
\begin{figure}[ht]
    \centering
    \includegraphics[width=\linewidth]{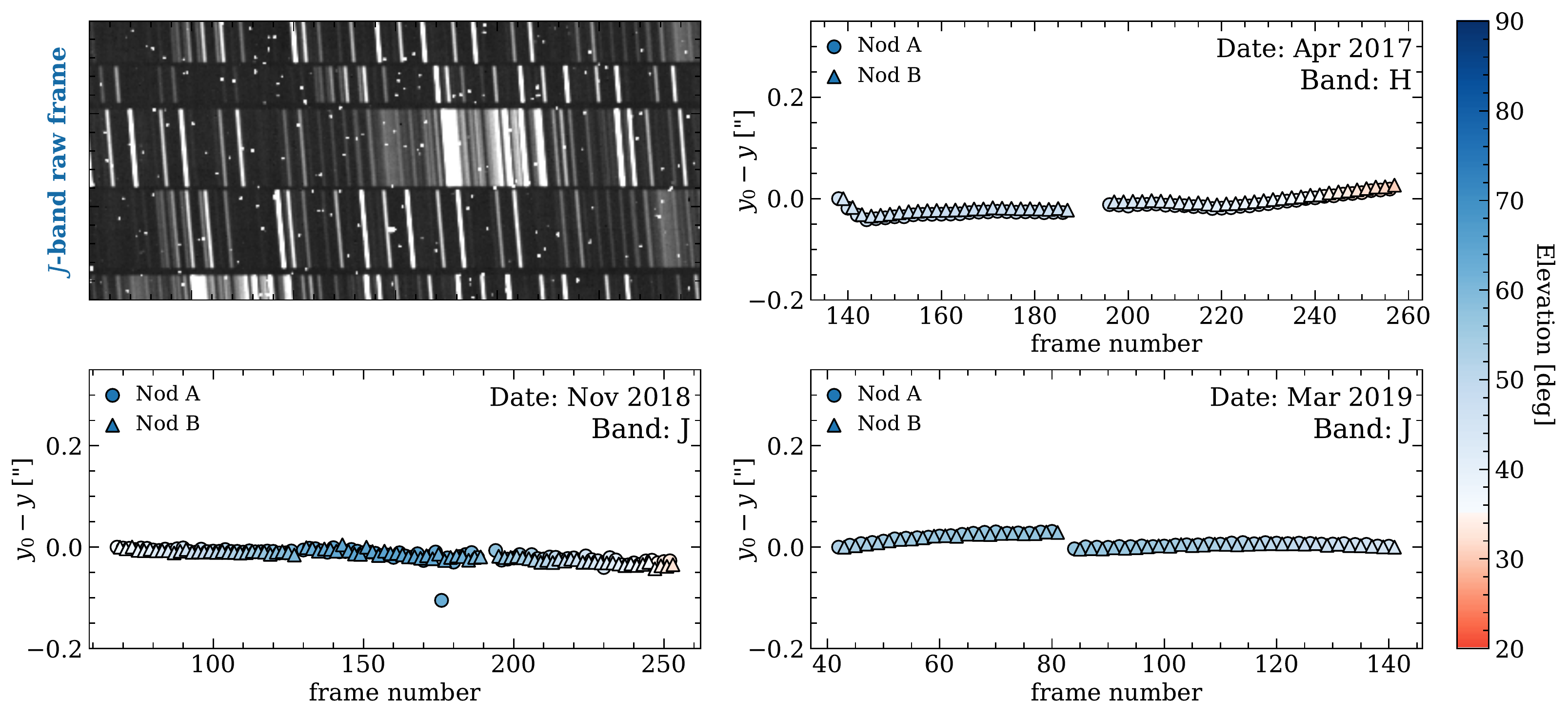}
    \caption{(\textit{top left}) An example of a raw $J$-band frame, zoomed in to shows a handful of slits in the data, highlighting how the atmospheric skylines can be arranged depending upon where the slits are located in the mask. (\textit{top right \& bottom}) The same datasets from Figure \ref{fig:star-drift} showing different examples of the (very small) relative mask drift in these datasets.}
    \label{fig:slit-drift}
\end{figure}

The method for measuring the shift of the atmospheric skylines in the raw frames has been briefly described in the beginning of \sref{ssec:insights}, however we will go into greater detail here.  In general, the raw MOSFIRE science mask is comprised of many slits, spaced vertically, and a forest of atmospheric skylines.  An example of a MOSFIRE $J$-band raw frame is shown in the top left panel of Figure \ref{fig:slit-drift}.  If the FCS was improperly accounting for the flexure at a given elevation or rotation angle of the telescope, then a cross-correlation between the first raw frame and the $n^{th}$ frame would show a significant shift in $x$ or $y$ (in image coordinates).  This shift would come from the warping of the data from the incorrect correction applied through the FCS mirror.  

For this cross-correlation, we used the \texttt{python} Image Registration package\footnote{Code available here: \href{https://github.com/keflavich/image_registration}{github.com/keflavich/image\_registration}} which works with extended emission, perfect for the skylines in the data (refer to Figure \ref{fig:slit-drift}).  The cross-correlation provided measured shifts in ($x$,$y$) coordinates, of which we focused on the $y$ shift.  Figure \ref{fig:slit-drift} shows the measured science mask shift for the same datasets shown in Figure \ref{fig:star-drift}.

The small drift apparent is well within accepted thresholds, therefore we infer that the remaining drift found by observers is likely not due to an improper correction of the internal flexure by the FCS.  However, we will detail the testing done to confirm this in \sref{ssec:test-fcs}.

\subsection{Testing FCS} \label{ssec:test-fcs}
The FCS testing was done using daytime engineering, as we could replace the atmospheric skylines with the lines from an Argon arc lamp and achieve the same results.  As described in \sref{sec:culprits}, the MOSFIRE spectrograph has a configurable slit mechanism that can allow for up to 46 slits in one mask (using a cryogenic robotic slit mask system, which can be changed electronically and reconfigure into a new mask in under five minutes).  In order to get the best measurements possible, we took advantage of all 46 slits and created a random pattern where each slit was in its own location horizontally.

The daytime testing was set up as follows:
~\vspace{-7mm} \\
\begin{enumerate}
    \item turn on the arc lamp
    \item turn off the FCS, choose a band for spectroscopy
    \item run through a large range of elevation: EL = (15\de $-$ 85\de$\!$) 
    \item at each elevation, take spectra at multiple rotation values from -360\de $-$ 45\de$\!$ 
    \item through all of this, verify that the FCS piston values are not changing (it should be off)
\end{enumerate}

\begin{figure}[ht]
\begin{minipage}{0.445\textwidth}
    \centering
    \includegraphics[trim=0 -10 0 0,clip,width=\linewidth]{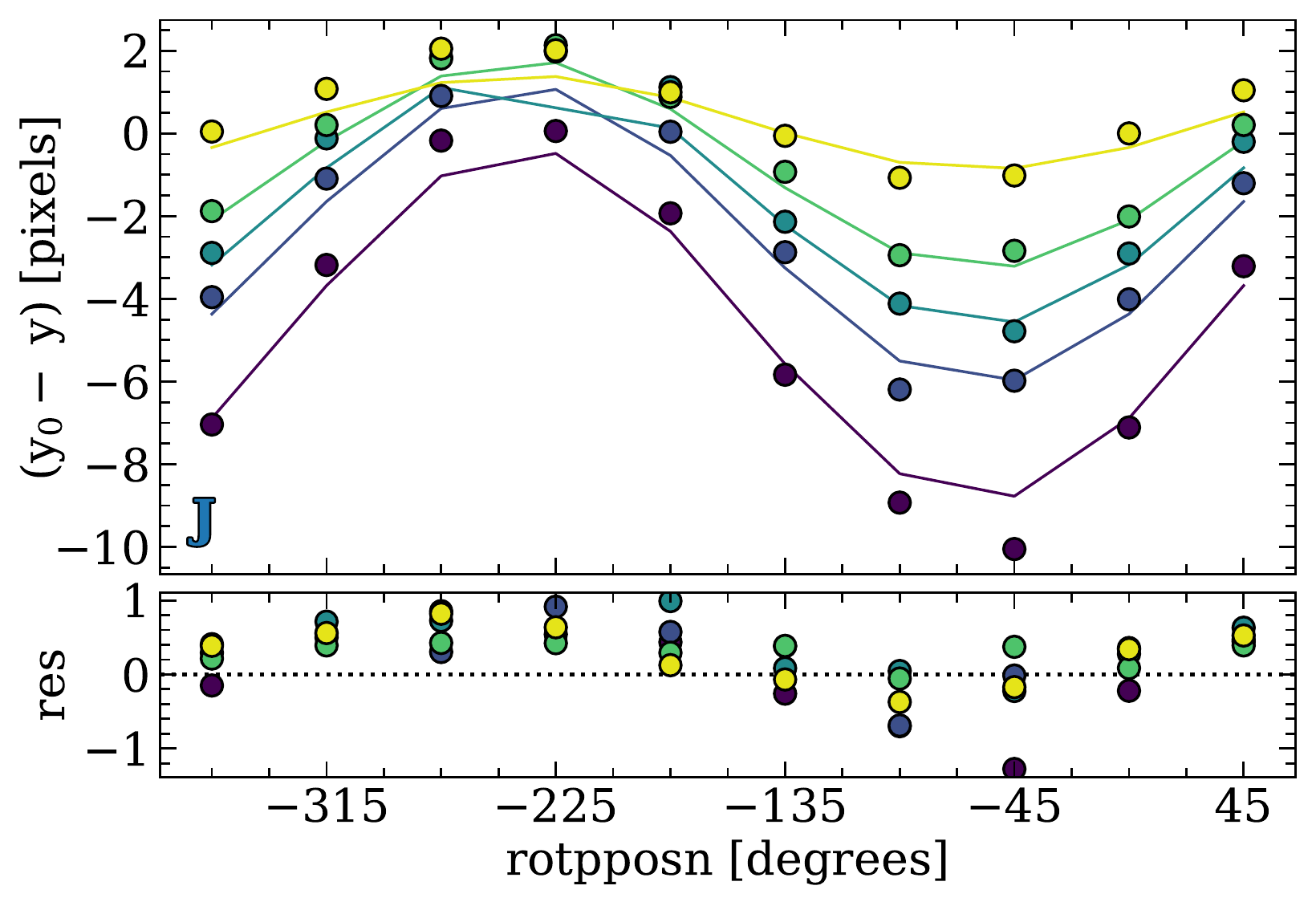}
\end{minipage}
\begin{minipage}{0.54\textwidth}
    \centering 
    \includegraphics[trim=0 -10 0 0,clip,width=\linewidth]{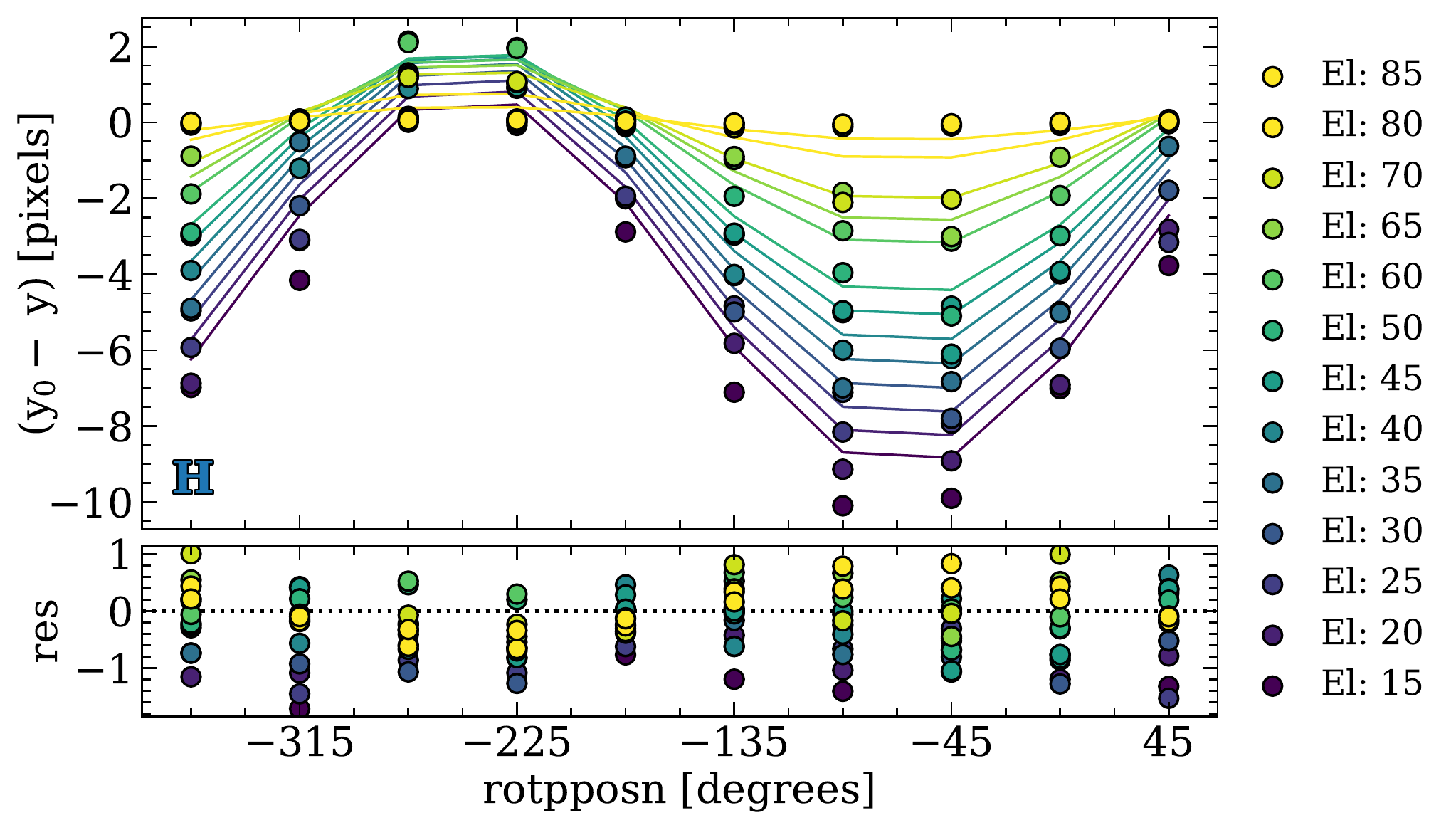}
\end{minipage}
\caption{\label{fig:fcs-off} Spatial offsets from the (\textit{left}) $J$-band and (\textit{right}) $H$-band testing, when the FCS was turned off.  The data shown were compared to a reference frame of EL = 45\de and \rotpposn = -90\de$\!$.  The lines overlaid are the FCS model fit to the data for each elevation, with the residual from the fits shown in the bottom panels.}
\end{figure}

We ran these tests on two different MOSFIRE bands ($J$ and $H$) in order to test both the $YJ$ grating and the $HK$ grating. Additionally, we ran the same tests again (in both bands) with the FCS on for a proper comparison. Using the Image Registration package, we measured the cross-correlation between each raw frame and a reference frame (chosen to be EL = 45\de and \rotpposn\,= -90\de$\!$, as described in \sref{ssec:fcs}) for each dataset.  The results from the tests with the FCS off are shown in Figure \ref{fig:fcs-off} for both bands, where the data are shown as the shift in y compared to the reference frame as a function of \rotpposn.

The scatter points in Figure \ref{fig:fcs-off} are colored to indicate elevation, and the colored lines overlaid are the fits to the FCS model used by MOSFIRE.  The residual of these fits are shown in the bottom panel of each plot, suggesting that the FCS does an acceptable job in correcting for the internal flexure over a wide range of elevation and rotation.  The model fits stray the most from the data points at the lowest elevations, but this is to be expected as the lowest elevations are a more extreme regime.

\begin{figure}[ht]
\begin{minipage}{0.46\textwidth}
    \centering
    \includegraphics[trim=0 -10 0 0,clip,width=\linewidth]{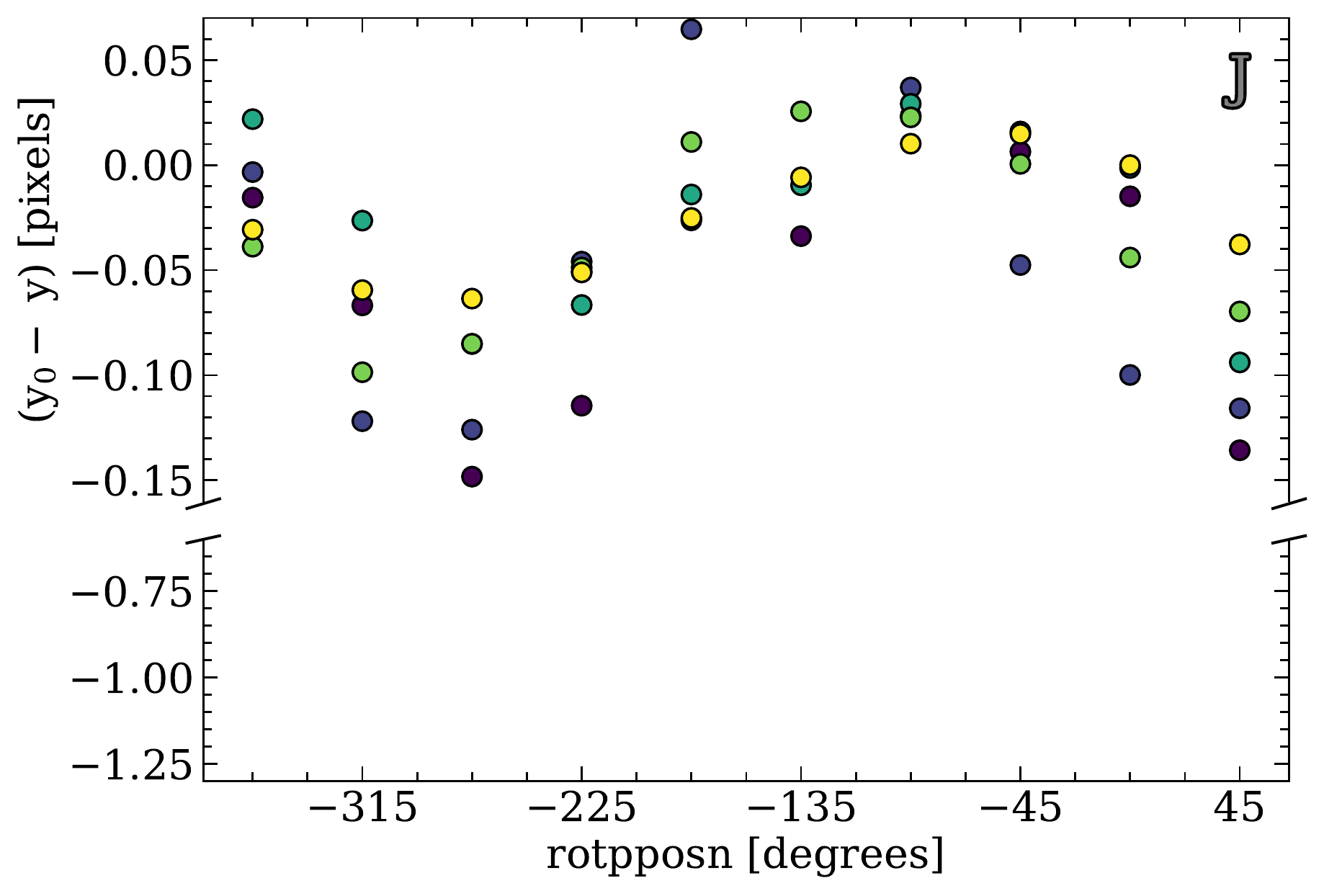}
\end{minipage}
\begin{minipage}{0.525\textwidth}
    \centering 
    \includegraphics[trim=0 -10 0 0,clip,width=\linewidth]{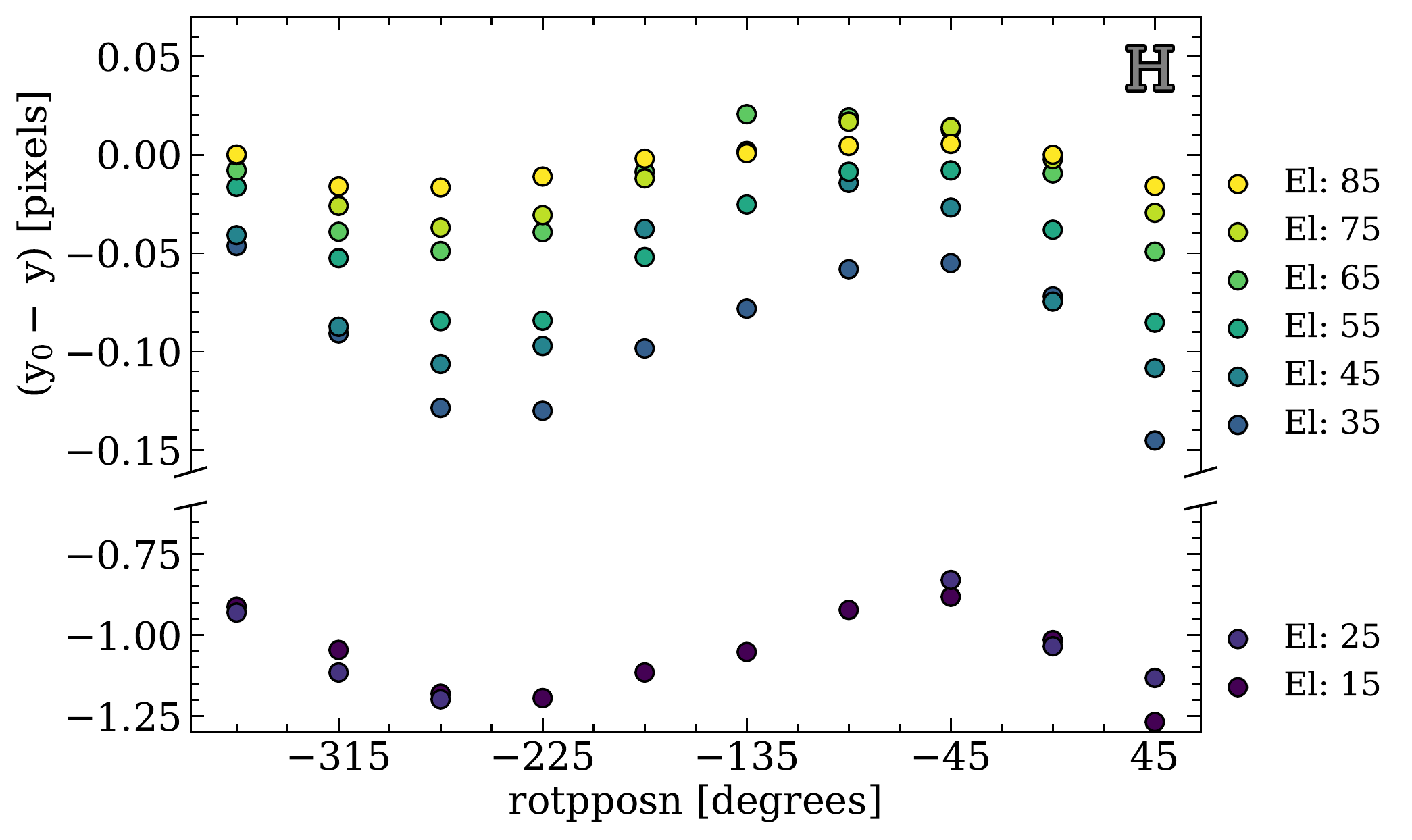}
\end{minipage}
\caption{\label{fig:fcs-on} Spatial offsets from the (\textit{left}) $J$-band and (\textit{right}) $H$-band testing, when the FCS was left on.  The data shown were compared to a reference frame of EL = 45\de and \rotpposn = -90\de$\!$.  Note that the spread of these curves is very small (with a small exception at the lowest elevations), indicating that the FCS is acceptably compensating for the changing internal flexure.}
\end{figure}

Next, we looked at the results from the same testing procedure, but with the FCS on, shown in Figure \ref{fig:fcs-on} for both bands.  The data are shown in the same format as Figure \ref{fig:fcs-off}; however, no line fits were applied as the FCS model was already in use (FCS was on).  The data are separated by elevation in the $H$-band plot, as the lowest elevations had larger spread.  Nevertheless, Figures \ref{fig:fcs-off} \& \ref{fig:fcs-on} clearly show the FCS is correcting for the changing internal flexure well within accepted values (e.g., the sub-pixel spread for EL $>$ 25\de$\!$).

\subsection{Testing Guider Flexure} \label{ssec:test-guider}
There are many ways one can test the flexure between the optical guider and the NIR science detector, using both spectroscopy and imaging.  As this flexure is the primary culprit based upon our insights from previous observational data (\sref{ssec:insights}), the majority of the focus for testing was done for this system. 
Thus far, we have run two different kinds of testing for this flexure, which we will outline below -- both of which require nighttime testing.

\hspace{3.65mm} \textbf{Imaging:}
~\vspace{-7mm} \\
\begin{itemize}
    \item at fixed elevation, focus on a guider star and rotate the detector \\
    in a great circle around that central point (the guider star)
\end{itemize}

\hspace{3.65mm} \textbf{Spectroscopy:}
~\vspace{-7mm} \\
\begin{itemize}
    \item changing guider flexure model coefficients
    \begin{enumerate} 
        \item make a science mask with bright targets, including many stars
        \item observe for $>$2 hrs on mask while tracking changes in drift
        \item modify the guider flexure model coefficients
        \item observe for $>$2 hrs on mask while tracking changes in drift
    \end{enumerate}
\end{itemize}

Below, we will outline the details for each test and share the results.

\begin{figure}[htb]
    \centering
    \includegraphics[width=0.9\linewidth]{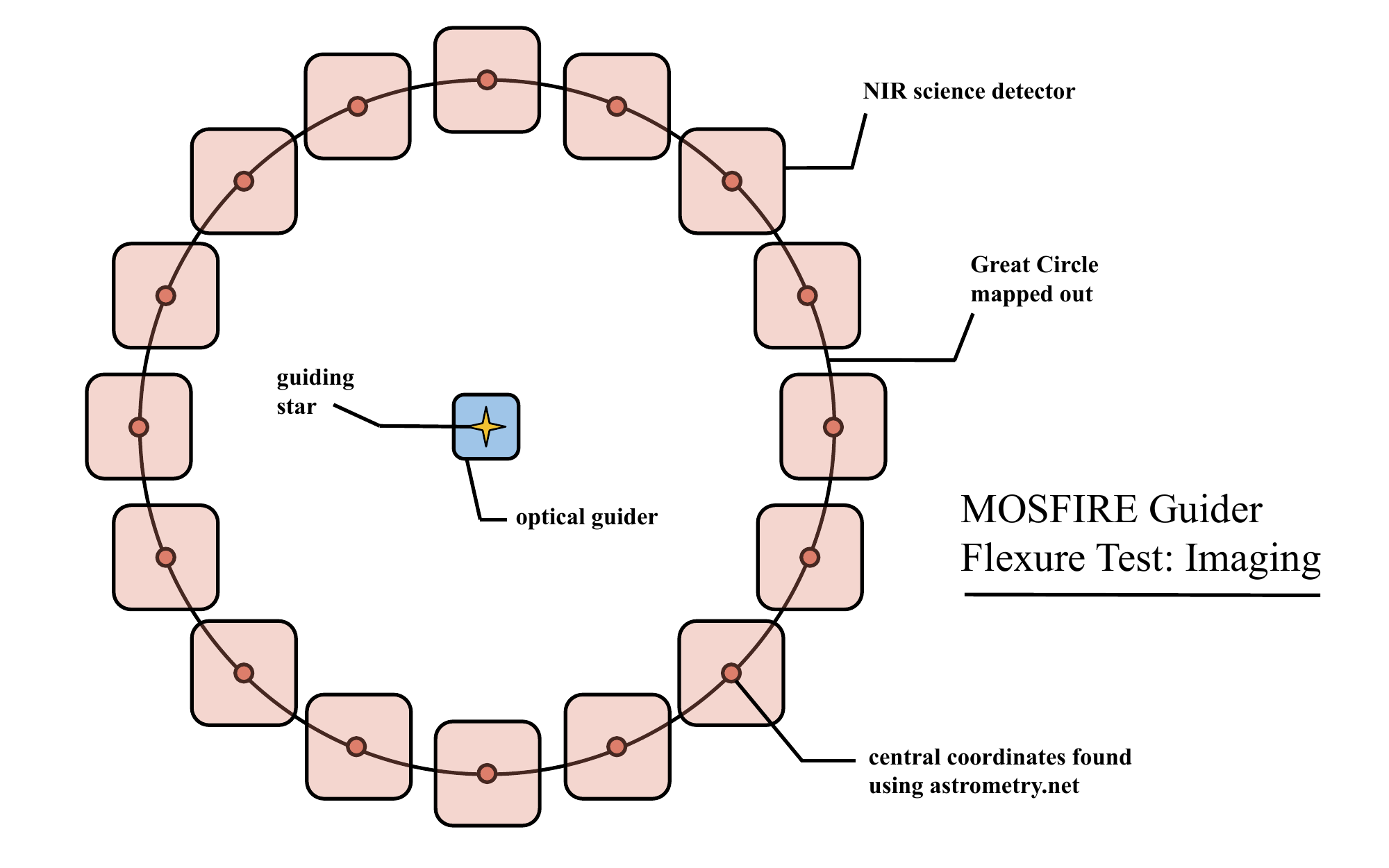}
    \caption{A basic schematic of the testing setup used for \sref{sssec:guider-fixed-el} (not to scale).  Using the same guiding star at a fixed elevation, and rotating the science detector around it, we ran imaging covering a full 360\de circle.  If there was no significant flexure compensation still needed between the optical guider and the NIR science detector, this test would be able to map out a near-perfect great circle on the sky.}
    \label{fig:great-circle}
\end{figure}

\subsubsection{Imaging: Fixed Elevation} \label{sssec:guider-fixed-el}
Choosing a field towards the South, as to avoid rapidly changing elevation, we had the Observing Assistant (OA) choose a pointing star near our desired elevation.  Beginning at a drive angle of -360\de, we centered on the guiding star and began imaging the sky where the science detector was pointed (offset from the guiding FOV).  Next, we rotated by 45\de and began the centering and imaging process again, continuing this until we rotated a full 360\de.  Finally, after a full rotation, we changed elevation and began again.  We chose to run this imaging in the MOSFIRE $Y$-band for the entire night, to minimize as much as possible the DAR between the optical guider and the MOSFIRE detector.

Figure \ref{fig:great-circle} shows a basic schematic of what this process looked like.  In theory, if the remaining flexure between the guider and the science detector was minimal, then the path mapped out by this process would be a great circle.  However, if the remaining flexure was significant, then there would be imperfections in this great circle.  In order to measure the center of each image, we used the \texttt{astrometry.net} software. The software produces central astrometric coordinates for a given field of stars, which we used to measure the center of the guider and science detector FOVs.  From there, we calculated the separation of the two FOVs and the physical angle between the guider and detector (after the angle of the detector has been subtracted off).  Both of these values should remain relatively constant if there is minimal remaining flexure. We converted these measurements to ($x,y$) coordinates and calculated the ($x$,$y$) offsets for each frame from where the expected center would be located for that frame.

\begin{figure}[ht]
    \centering
    \includegraphics[width=0.9\textwidth]{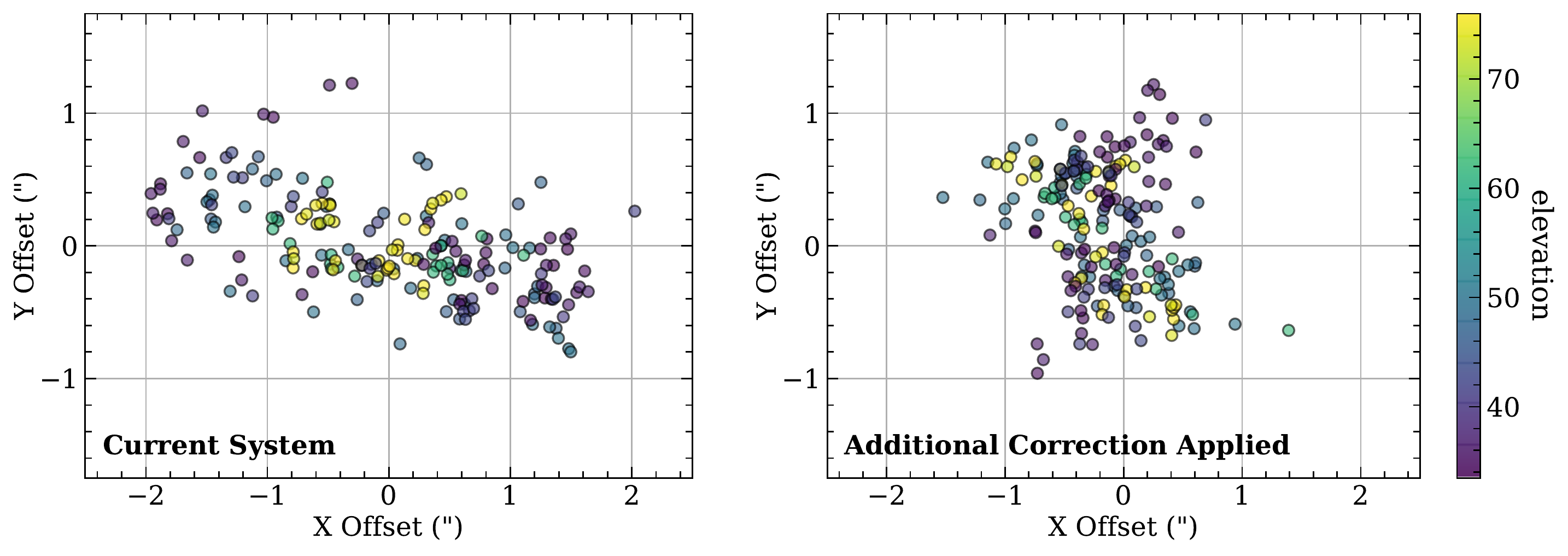}
    \caption{Coordinate offsets in the guider FOV for the rotating testing, colored by elevation. (\textit{left}) Offsets from testing using the current system.  (\textit{right}) The same offsets with additional correction applied, shrinking the spread of points. }
    \label{fig:guider-correction}
    ~\vspace{-6mm}\\
\end{figure}

Figure \ref{fig:guider-correction} shows all measured offsets in ($x,y$) space.  The left panel shows the spread of measurements from the data with no additional correction applied, therefore only the current guider flexure corrections are in use.  As a reminder, Figure \ref{fig:guider-optics} shows the coordinate layout for the guider system compared to the science detector -- in this system, the $x$ direction in the guider FOV corresponds to the $y$ direction in the science detector's FOV (the spatial direction). The right panel shows these data with an additional correction applied, with $x$ and $y$ models created using a sinusoidal combination of elevation and rotation.  Note that even with this additional correction applied, there remains some shape to the distribution. Further testing covering a larger range of elevation and location on the sky is needed to better approximate the model for use in this correction. 
\note{However, we tentatively conclude that there exists a hysteresis component which is uncorrected by the current guider flexure model.}

\subsubsection{Spectroscopy: Taking Data with Modified Flexure Model}  \label{sssec:guider-spec} 
As briefly described in \sref{ssec:test-guider}, this test used spectroscopy with a full mask of targets -- where several of these targets were faint stars (16 mag $< H_{160} <$ 19 mag) that would be easily detectable in the raw frames.  The setup of this test was straightforward: take observations on the mask using the original guider flexure model parameters, modify the coefficients used in the guider flexure model and take observations, modify the coefficients further and take more observations.  While the data were coming in, we tracked the drift of the stars in the mask using the techniques described in \sref{sssec:insights-star}. Throughout these observations, we did not stop to realign as we were specifically looking to measure the breadth of the drift in each dataset. 

\begin{figure}[ht]
    \centering
    \includegraphics[width=\linewidth]{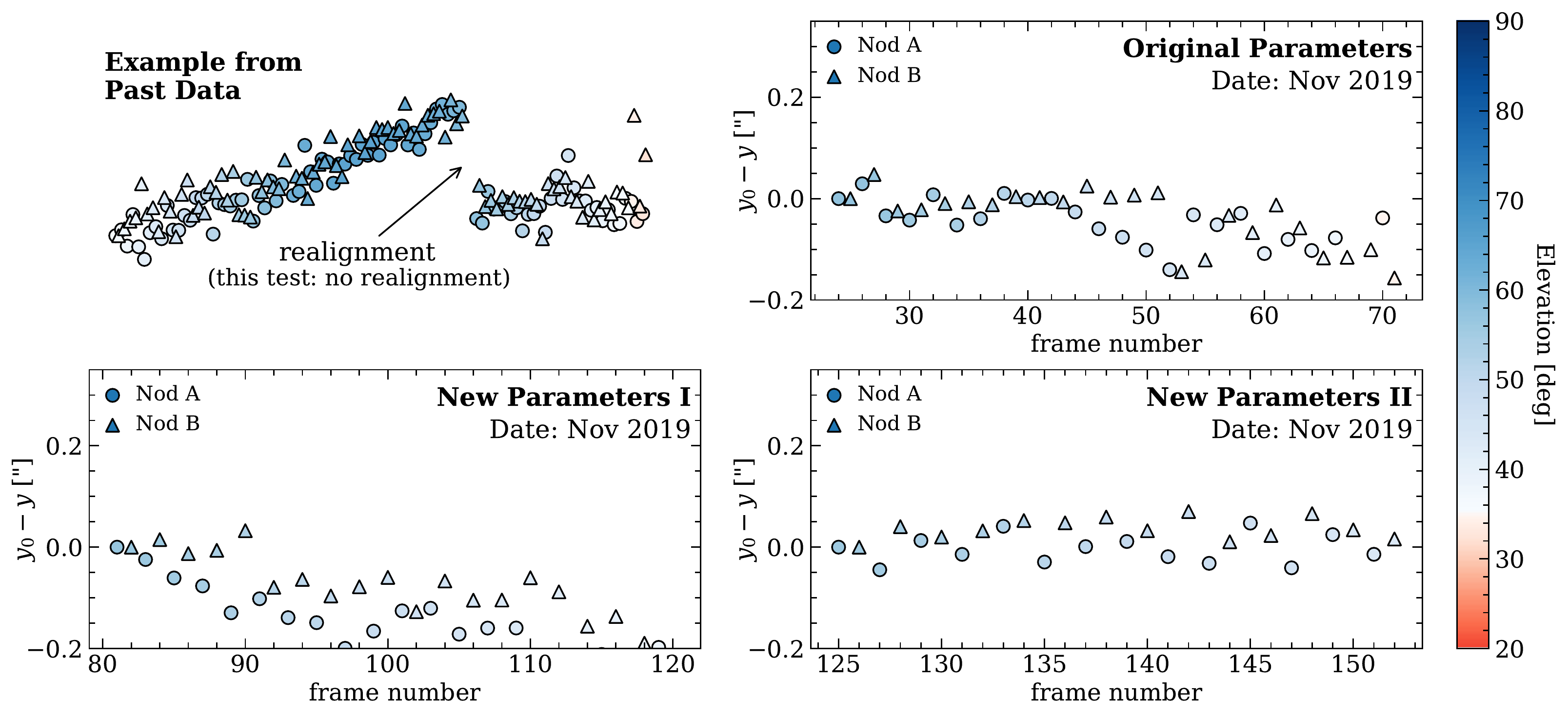}
    \caption{(\textit{top left}) An example of the star drift measured in previous MOSFIRE data (these data are also shown in Figure \ref{fig:star-drift}), with the moment of realignment highlighted; for this test we purposefully did not realign. (\textit{top right \& bottom}) The data resulting from modifying the guider flexure model parameters.}
    \label{fig:guider-test-star}
\end{figure}

Figure \ref{fig:guider-test-star} shows the three preliminary datasets created from this testing process.  Most notably, the final coefficients used in the guider flexure model (bottom right) appear to reduce the remaining drift the most -- however more testing is needed to truly confirm this.  Further testing would cover a larger range of field locations on the night sky, as well as a larger range of elevation.

\subsection{Testing DAR} \label{ssec:test-dar}
The testing planned to check the remaining effect of the DAR have yet to be completed, however here we will outline the steps that will be taken.  Broadly, the MOSFIRE guiding system (MAGIQ) has a series of equations used to correct for the DAR given the optical guider's bandpass, the science detector's bandpass (chosen by the user), and the elevation of the telescope.  Using the \texttt{GUIDWAVE} and \texttt{TARGWAVE} keywords for the target wavelengths for the guider and science detectors, respectively, one can calculate the magnitude of the DAR correction through a range of elevations.

Corrections were calculated and applied in 2015\cite{kass15} to account for the half-magnitude effect they had measured in the drift, therefore we find it unlikely that the DAR is a real culprit in this remaining drift. However, we commit to conducting this testing to answer this question conclusively.

\section{Discussion \& Future Work} \label{sec:future}
From the testing done to date, we conclude that the remaining drift measured by observers is likely due primarily to the remaining uncorrected flexure between the optical guider and the NIR science detector.  This flexure effect has been well-modeled in the past\cite{kass15}, however we aim to increase the precision further such that the lingering drift remains sub-pixel throughout even the lengthiest of observations.

Through testing the FCS, we found that it is appropriately compensating for the internal flexure experienced by the MOSFIRE spectrograph through a wide range of elevation and rotation (\rotpposn).  We note that further testing and refinement could prove useful for the lowest elevations (EL $<$ 20\de), although this is not a dire need as most observers never push to those depths when tracking cosmic objects. 

We have plans to complete testing for the DAR in order to confirm whether or not there is a residual compensation still needed, despite our suspicions that there will be none.

\subsection{Future Testing} \label{ssec:future-testing}
The testing planned for the DAR has been described in \sref{ssec:test-dar}, however additional testing could be done using imaging of bright stars and comparing their locations in the guider FOV versus the science detector's FOV. 
Additional testing will need to be done on the residual flexure component between the optical guider and the NIR science detector.  This testing will allow for a more complete analysis of the current model used and of the ways that its coefficients can be modified to better account for the remaining drift seen.  Further tests that can be done to enable this effort include more spectroscopic observations (as described in \ref{sssec:guider-spec}) where the guider coefficients are tuned; replacing alignment procedures during spectroscopic testing with imaging (to locate where the alignment stars are in their box slits); and running more imaging campaigns (focused on mapping out great circles around a central guiding star) at more elevations.

Another possible test that can be used to investigate the remaining flexure between the optical guider and the NIR science detector could be in tracking a rising star field.  This test is briefly described below.

\subsubsection{Guider Testing: Tracking a Star Field} \label{sssec:guider-star-field}
\begin{figure}[ht]
    \centering
    \includegraphics[trim=0 100 0 31,clip,width=0.85\linewidth]{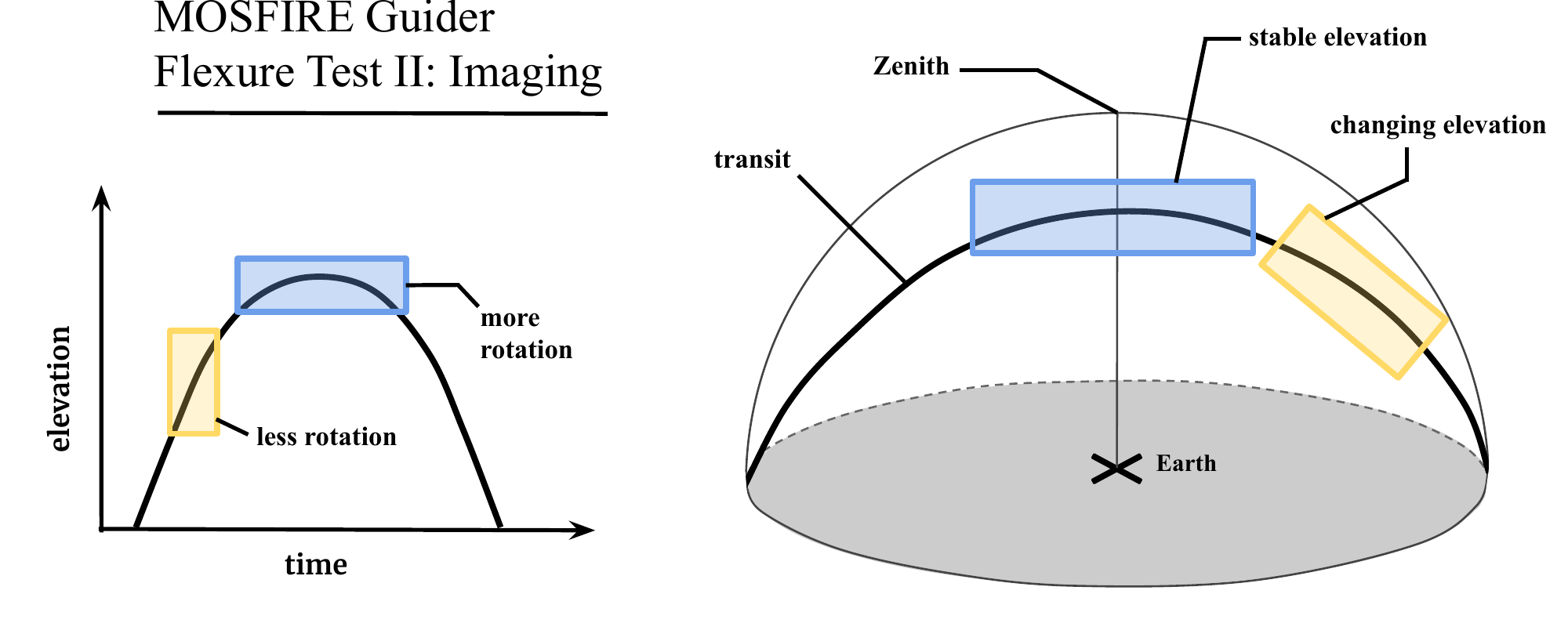}
    \caption{A basic schematic showing the two different tracking testing that can be done using a single rising star field (\sref{sssec:guider-star-field}).  As the star field begins to rise in the night (yellow box), the elevation of the field changes rapidly but the rotation of the detector would be minimal. As the star field traverses its highest point in transit (blue box), the elevation of the field would remain relatively constant while the rotation of the detector would be changing more rapidly.}
    \label{fig:changing-el}
\end{figure}

The goal of this investigation would be to take advantage of the different testing that can be done with tracking a single star field over the course of the night.  This test would allow us to 1) follow the star field when it was rising, therefore changing in elevation rapidly but rotating less, and 2) track the star field as it transits close to zenith, where there is minimal change in elevation but rapid rotation. Figure \ref{fig:changing-el} shows a basic schematic for how this testing would work, highlighting the areas in the field's path where is it rising quickly (gold boxes) and where it is passing the peak of its transit (blue boxes).

\section{Conclusion}

We have presented a recent (and ongoing) evaluation of the Multi-Object Spectrometer For Infra-Red Exploration (MOSFIRE) on the Keck I telescope; work done in response to a spatial drift measured in spectroscopic data by MOSFIRE users.  While the measured drift of 0.5 pix/hr does not impact the science done by most MOSFIRE users, this effect can be significantly detrimental to users whose science requires lengthy observations of faint sources for more than four hours on one science mask.  This niche group of users, albeit small, represent some of the most groundbreaking work that can be done using MOSFIRE. 

We identify the measured drift to be the possible result of three factors: the internal flexure compensation system (FCS), the optical guider camera flexure system, and/or the differential atmospheric refraction (DAR) corrections.  In this work, we described the three systems in detail and walked through the various testing completed, highlighting where further testing is needed.  In summary, we pinpoint the flexure between the optical guider and the NIR science detector as the most likely contributor to the remaining measured drift.  Further testing is needed to modify the current flexure model accurately and precisely to correct for the drift.


\acknowledgments     
 
The authors wish to recognize and acknowledge the very significant cultural role and reverence that the summit of Mauna Kea has always had within the indigenous Hawai\,$\!`$ian 
community. We are most fortunate to have the opportunity to conduct observations from this sacred mountain. 
The authors thank P.\ Gomez for helpful insight. The authors thank the W.\ M.\ Keck Observatory OAs who enabled the many hours of detailed testing done thus far. TAH acknowledges generous support from the Texas A\&M University and the George P. and Cynthia Woods Institute for Fundamental Physics and Astronomy, which resides on indigenous Sana and Tonkawa land.\footnote{This information comes from the Canadian non-profit \href{native-lands.ca}{native-lands.ca} database via \href{land.codeforanchorage.org}{land.codeforanchorage.org}} 
This material is based upon work supported by the National Science Foundation Graduate Research Fellowship.

\bibliography{main} 
\bibliographystyle{spiebib} 

\end{document}